# Methodological reflections for AI alignment research using human feedback

Thilo Hagendorff
thilo.hagendorff@uni-tuebingen.de
University of Tuebingen

Sarah Fabi
sfabi@ucsd.edu
University of California San Diego

**Abstract** – The field of artificial intelligence (AI) alignment aims to investigate whether AI technologies align with human interests and values and function in a safe and ethical manner. AI alignment is particularly relevant for large language models (LLMs), which have the potential to exhibit unintended behavior due to their ability to learn and adapt in ways that are difficult to predict. In this paper, we discuss methodological challenges for the alignment problem specifically in the context of LLMs trained to summarize texts. In particular, we focus on methods for collecting reliable human feedback on summaries to train a reward model which in turn improves the summarization model. We conclude by suggesting specific improvements in the experimental design of alignment studies for LLMs' summarization capabilities.

**Keywords** – AI alignment, scalable alignment, summarization, large language models

## 1 Introduction

Since artificial intelligence (AI) technologies are becoming increasingly pervasive in many fields of society, it is important to investigate whether they align with human interests and values, and whether they function in a safe and ethical manner. AI alignment is the field of study concerned with that (Gabriel 2020). One of the key challenges in AI alignment is that it is difficult to anticipate how an AI system will behave once it is deployed, as it may be able to learn and adapt in ways that are difficult to predict (Wei et al. 2022). This can lead to unintended consequences, such as an AI system making decisions that are harmful to humans or animals. To address these challenges, researchers in the field of AI alignment are working on a variety of approaches, including designing AI systems with built-in ethical constraints (Jiang et al. 2021), developing methods for verifying and validating the behavior of AI systems (Rahwan et al. 2019), and creating ways to ensure that AI systems can be controlled and turned off if necessary (Amodei et al. 2017).

The motivation for AI alignment has its roots in fears of malevolent traits of artificial general intelligence (AGI) (Bostrom 2014). However, since the potential advent of AGI remains speculative (Floridi 2022), most current alignment research is more focused on tangible problems with existing AI technologies, especially large language models (LLMs). Alignment problems with LLMs can be investigated by evaluating their outputs considering human values and interests (Weidinger et al. 2021). However, these values can be vague or even unknown, and still, LLMs should not violate them. This so-called "scalable alignment" problem can be investigated in several regards. Let's consider the following example: An LLM is trained to produce summaries of longer texts. Here, values or criteria for a good summarization may in part be unknown, but LLMs should nevertheless be able to stick to them. This problem and the corresponding human feedback loops can be a prototype for alignment problems with LLMs in general, meaning that the methodological considerations hold for other fields of alignment, too.

Hence, in this paper, we discuss the methodology of the alignment problem regarding summarization capabilities of LLMs. We start from the premise that a typical setup is used (Stiennon et al. 2020; Ziegler et al. 2020): AI researchers train a summarization model with the help of a reward model, which had been trained on human evaluators' ratings of existing summaries. We further assume that the fine-tuned summarization model generates summaries that are rated higher than the initial summaries by a new group of human evaluators. In this setting, all models and human evaluators are presented with the original texts as well.

In the following, we first perform some general analyses to get a better understanding of the specific alignment problem in this setup. Secondly, we introduce several techniques to tackle the problem and present a brief overview of potential amendments to alignment methods in the summarization problem.

## 2  Problem analysis

In the first step of tackling the alignment problem for summarization capabilities in LLMs (Stiennon et al. 2020; Ziegler et al. 2020), human feedback from AI trainers is collected to assess a data set of existing summaries. Based on patterns of this human feedback, a reward model is trained. This model is then deployed in a reinforcement learning setting to update the summarization capabilities of the LLM. This way, human feedback is automated to scale it up so that a plethora of synthetic training stimuli which are aligned with human values can be generated and used to improve the LLM's summarization capabilities. However, this approach still faces methodological challenges that can thwart successful alignment. These challenges can emerge in several areas that we discuss in the next sections, for instance when AI researchers assess the summaries and feel that they are biased, error-prone, or possess other shortcomings.

### 2.1  Error-proneness

A potential problem can occur when AI researchers find the AI-generated summaries to be erroneous. A likely possibility for this is that before training the reward model, AI trainers oversaw errors in the summary data set due to a lack of motivation. This could have many reasons like time pressure when being paid by summary. Moreover, it is important to check whether the summaries in the original data set compressed the information to such an extent that they seem to be easily understandable for non-experts like AI trainers, whereas the AI researcher found this information to be oversimplified and false. If data sets consist for instance of paper summaries (like SCITDLR (Cachola et al. 2020) or Multi-XScience (Lu et al. 2020)), this makes the judgment of summaries hard for non-scientists. To address this problem, it would be helpful to apply "sandwiching techniques" (Cotra 2021) which aim at closing the gap between expert and non-expert feedback. This will be discussed in more detail in 2.4.

## 2.2 Biases

A further problem can occur when AI researchers find the AI-generated summaries to be biased. These biases can be rather harmless, for instance when addressing the heading bias where people mostly rely on the title of a text for summarization (Lorch et al. 2001). On the other hand, social biases in summarizations can stand in direct conflict with alignment goals. In our setting, the summaries could be biased because of biases already existing in the original texts. When the texts contain factual information, it is more probable that the bias originates from the AI trainers instead of the texts. AI trainers as well as researchers are biased by their demographical background, leading to specific topics being represented more frequently in the summaries because of different individually perceived relevance (Hahn and Garner 1985; Hidi and Anderson 1986). Therefore, one has to look at the demographical background of the AI trainers to assess their potential biases. Assuming one has the same demographical data as for instance Stiennon et al. (2020) in their study on learning to summarize from human feedback – meaning mostly young individuals (20-29) and binary genders – one can assume that information that is of interest for elderly people or people of nonbinary gender is underrepresented in the final summaries.

## 2.3 Verbosity and copying

Furthermore, to identify whether AI trainers (erroneously) prefer longer texts, one can look at the correlation between the word count of the summaries and the corresponding ratings. AI trainers might have focused on the completeness of the information in the summary, whereas they did not evaluate how well-compressed the information was. This might have influenced the reward model's training. Another methodological problem occurred in Ziegler et al. (2020). The researchers found their system to be extractive instead of abstractive, meaning that the summaries consisted of large parts of copied original text. This problem can be detected by looking for the longest common subsequences of the original text and its summary.

## 2.4 Agreement

In order to investigate potential disagreements between AI researchers and trainers in more detail, one could invite further researchers to rate those summaries which have already been labeled by various AI trainers (for example in Ziegler et al. (2020), 5 % of the summaries have been rated not by one but five AI trainers). With this new data, one could look at the researcher-trainer agreement (as the mean squared difference of the labels over all summaries) to rule out that it was just the one AI researcher that disagreed with the trainers. Furthermore, one could calculate the researcher-researcher agreement. Previous experiments show that researchers do not necessarily agree on how summaries should be judged: In Ziegler et al. (2020), researchers only agree in 60 %, and in Stiennon et al. (2020) in 73 % of cases. The reported result of a 95 % agreement rate after discussion in Stiennon et al. (2020) should be perceived with caution. Psychological studies that investigate how groups come to agreements after discussions show that these agreements are not necessarily better than previous judgments due to humans' desire for conformity (Sherif 1936). Thus, researchers seem to be experts but could also use some enhancement to serve as a good baseline. The trainer-trainer agreement is also interesting to look at since it provides information about how consistent the ratings are across trainers. There might even exist a subgroup whose judgments are closer to the researchers', possibly people with strong language skills or scientific background. Further analyses on the ratings could be performed, for instance to confirm a wide range of different quality levels of the summaries. Just to be sure, AI researchers could also look at the agreement between the reward model and the AI trainers.

## 3 Improvements in the experimental design

After identifying specific difficulties in the sections above, we want to describe specific suggestions for improving the experimental design of alignment studies using human feedback.

### 3.1 Evaluation criteria

How researchers communicate to AI trainers about criteria for high-quality summaries is crucial for alignment studies to work. In the present study design, it might have been the case that the AI trainers did not know what exactly they should focus on during their rating. Researchers should agree upon several criteria according to which the summaries should be judged. Inspired by the criteria used to judge ML models' summaries (Huang et al. 2020; Wu et al. 2021), as well as criteria from linguistics to judge human summaries (Hidi and Anderson 1986; Soubbotin and Soubbotin 2015), we suggest the following criteria: Accuracy, coverage, condensation, and fluency. By adding the criterion condensation, we hope to circumvent the problem of Stiennon et al. (2020) who reported length to be a significant confounder of quality. Alternatively, one could control for length or ask AI trainers 'how good is the summary, given that it is X words long?' to avoid models generating the longest summaries allowed by the length constraints (Wu et al. 2021). Based on the four criteria, we suggest building a framework providing the labeling researchers and trainers with exemplary questions with which they can address each criterion. This framework can be downloaded in the beginning so that it is accessible during the whole labeling process. Moreover, based on the criteria, AI trainers should rate the quality of the summary on a 6-point Likert scale ('very bad', 'bad', 'a bit flawed', 'ok', 'good', 'very good') to not give too many options, making answers more consistent and to not provide a medium label (psychological research shows that when providing a medium score participants tend to use this more frequently). Labels will have to be normalized to account for participants who differ by a monotonic transform.

### 3.2 Instructions

AI trainers have to receive instructions. The phrasing of these instructions should be as precise as possible without making them too complex or too long. Instructions should be easily understandable and memorizable. Furthermore, AI trainers should be asked to focus on their own biases. We suggest explaining to them that primacy and recency effects might determine which parts of the original text they tend to remember. Moreover, we suggest pointing out that the proportions of content from different groups in the summary should resemble the proportions in the original text (proportional representation (Shandilya et al.

2020)). AI trainers should be asked to read the texts thoroughly without skipping parts of them and could furthermore be able to ask questions to AI researchers to make sure they understood the task and the criteria.

## 3.3 Practice trials

In psychology, for instance, practice trials (after the instructions, before the start of the actual experiment) are part of good scientific practice. Study participants get used to the task and can ask for clarification if they encounter problems. In our case, AI trainers could get used to the qualitative and quantitative dimensions of the summaries. This prevents later shifts in their rating technique. Furthermore, we would design practice trials that contain summaries that are especially tricky and address the problems that have been determined by our primary analyses, like verbose language, merely extractive or too simplistic summaries, and biases. After each practice trial, feedback should be provided to drive AI trainers' attention to the problematic areas. They will only be able to start the actual experiment after having rated at least 10 trials adequately. A difference of one point in the six-point rating scale compared to the expert ratings could be tolerated.

## 3.4 Quality checks

A further improvement could be to provide pre-rated summaries from time to time as quality checks. For these, AI trainers will receive feedback. If the answers on these quality checks of individual AI trainers deviate too much from the expert ratings, the corresponding data will not be used to train the reward model. Response times will also be monitored to make it easier to identify AI trainers who do not read the texts thoroughly. Another quality check could be that AI trainers rate how sure they are about their rating, even though in Stiennon et al. (2020), sorting out the trials with low confidence did not improve results. Wu et al. (2021) even used a mechanism where a second labeler provided feedback on task completion.

## 3.5 AI trainers

To ensure high labeling quality and high motivation, one should prefer AI trainers who were not recruited from a crowdsourcing platform but who have proficiency in language and writing – e.g., employees of media departments, similar to Wu et al. (2021). We suggest favoring having not too many but rather well-selected AI trainers that label lots of summaries. If this is too expensive, one can nevertheless select AI trainers with high language skills via a first screening test. It will be impossible to recruit AI trainers of all ethnicities, gender, age groups, etc. but since demographic data is collected one can at least identify potential sources of biases. Furthermore, we suggest trying to enhance the AI trainers' motivation. According to Rogstadius et al. (2011), higher payment does not necessarily improve motivation, whereas addressing their intrinsic motivation, for example by informing them about their important contribution to alignment research, is advantageous. Furthermore, AI trainers' performance can be improved by designing the user interface in an appealing way, maybe even by playing some music (Sampath et al. 2014).

## 3.6 Expert baseline

A further suggestion is to use handpicked experts who provide baseline labels. They will receive the same training as usual AI trainers. If the factual information does require expertise in a specific field, like computer science texts, researchers in the field will represent a good baseline. If no qualified expertise is required regarding the text content, it makes sense to select language experts. Again, instructing experts regarding biases and training them on the selected criteria will be important. We do not agree with Stiennon et al. (2020) who claim that only for more complex and less straightforward tasks one should consider the groups of individuals whose labels serve as a baseline, since summarizing machines might have a tremendous influence on our view of the world in the future and it would be problematic if certain groups would not be part of such summaries.

## 3.7 Sandwiching techniques

If after all of these measures the fact remains that ratings of non-experts and experts differ, for example, because the texts contain factual information about computer science topics, it will be helpful to look for so-called "sandwiching techniques" that help non-experts to provide labels that are very similar to experts' labels (Cotra 2021). We briefly present several approaches: If the texts are very long, it might be helpful to decompose the problem into subproblems and split it up among many AI trainers, starting with shorter paragraphs leading to longer text parts in a stepwise manner. A procedure to recursively summarize books with human feedback by decomposing the books' texts can be found in Wu et al. (2021). If the original texts contain lots of technical terms, on the other hand, one could train a model to explain those terms to AI trainers (Cotra 2021). An even more advanced approach would be to use an LLM to enhance human judgments directly. A very recent example can be found in Saunders et al. (2022), who trained a model to point out flaws in the summaries to enhance human judgments. Using this approach, the ratings are 'corrected' in one direction only (AI trainers find more critiques) because the model draws attention to the negative aspects of the summary. One could go one step further and try to not only eliminate flaws but also reinforce valuable summarizing behavior. Therefore, we suggest letting AI trainers observe two models debating about the value of the summary (Irving and Askell 2019; Irving et al. 2018). One debater could be trained with human texts critiquing summaries (like in Saunders (2022)), whereas the other debater would be trained with summaries and corresponding human-written positive feedback (again humans writing the positive and negative feedback have to be thoroughly selected and trained). The actual experiment would have the same structure as our original experiment except for the enhancement of the human feedback with two LLMs debating about the presented summary. Depending on the goal of the study, this approach might be a bit overpowered for the summarization task and not very time-efficient. If, on the other hand, we want to gain valuable insights for future, more complex tasks, it might be worth the effort.

## 4 Conclusion

Gaining insights into how we can enhance human feedback for tasks in which the AI models' capabilities exceed those of humans will be crucial for scaling AI alignment in the future and will go far beyond the scope of summarization tasks. In order to have a sound methodology for the latter, we suggest starting by selecting experts, fine-tuning the criteria, the corresponding framework, and the training to such an extent that the expert-expert agreement will reach a certain threshold. Only then can one be satisfied with using the experts' labels as a baseline. Then, a small subset of AI trainers should be invited to label a subset of summaries. If the expert-trainer agreement is not satisfying, we suggest introducing further sandwiching techniques.

Assuming that the texts are not especially long, we suggest starting with a model explaining potential technical terms to AI trainers. If this still does not enhance the expert-trainer agreement to the same level as the expert-expert agreement, we would consider giving humans feedback from an assisting LLM that is able to critique summaries. Only when we are satisfied with the expert-trainer agreement for this small subgroup, one should recruit and train more AI trainers with the same measures and let them take part in the whole experiment. We propose monitoring whether one must exclude AI trainers (see quality checks) and use the data to train the reward model and the summarization model in a semi-online manner. New AI trainers who rate the final summaries should be recruited and trained in the same manner. These steps ensure a sound methodology when using human feedback for AI alignment tasks, in particular with respect to summarization capabilities in LLMs.

## Publication bibliography